\pdfoutput=1

\documentclass[11pt]{article}

\usepackage[preprint]{acl}

\usepackage{times}
\usepackage{latexsym}

\usepackage[T1]{fontenc}

\usepackage[utf8]{inputenc}

\usepackage{microtype}

\usepackage{inconsolata}

\usepackage{graphicx}

\title{Plan with Code: Comparing approaches for robust NL to DSL generation}

\author{Nastaran Bassamzadeh\qquad Chhaya Methani\\Microsoft Corporation\\Redmond, USA}

\begin{document}
\maketitle
\begin{abstract}
Planning in code is considered a more reliable approach for many orchestration tasks. This is because code is more tractable than steps generated via Natural Language and make it easy to support more complex sequences by abstracting deterministic logic into functions. It also allows spotting issues with incorrect function names with the help of parsing checks that can be run on code. Progress in Code Generation methodologies, however, remains limited to general-purpose languages like C, C++, and Python. LLMs continue to face challenges with custom function names in Domain Specific Languages or DSLs, leading to higher hallucination rates and syntax errors. This is more common for custom function names, that are typically part of the plan. Moreover, keeping LLMs up-to-date with newer function names is an issue. This poses a challenge for scenarios like task planning over a large number of APIs, since the plan is represented as a DSL having custom API names. In this paper, we focus on workflow automation in RPA (Robotic Process Automation) domain as a special case of task planning. We present optimizations for using Retrieval Augmented Generation (or RAG) with LLMs for DSL generation along with an ablation study comparing these strategies with a fine-tuned model. Our results showed that the fine-tuned model scored the best on code similarity metric. However, with our optimizations, RAG approach is able to match the quality for in-domain API names in the test set. Additionally, it offers significant advantage for out-of-domain or unseen API names, outperforming Fine-Tuned model on similarity metric by 7 pts.
\end{abstract}

\section{Introduction}

\label{intro}
There has been significant progress made in improving and quantifying the quality of Natural Language to Code Generation or NL2Code (\citealp{Chen21CodeGenEval}, \citealp{Nguyen22Eval}). Recent improvements in models for general-purpose languages like Python, C++ and Java can be attributed to larger LLMs (\citealp{ChatGPT22}, \citealp{GPT423}) and the availability of Pre-trained open-source models (\citealp{CodeLlama23}, \citealp{abdin2024phi3}, \citealp{Codestral24}) advancing the state-of-the-art. However, there hasn’t been a focus on improving quality of Natural Language to Domain Specific Languages or NL2DSL, which a lot of enterprise applications rely on. 

\begin{figure*}
  \centering
  \includegraphics[width=\textwidth]{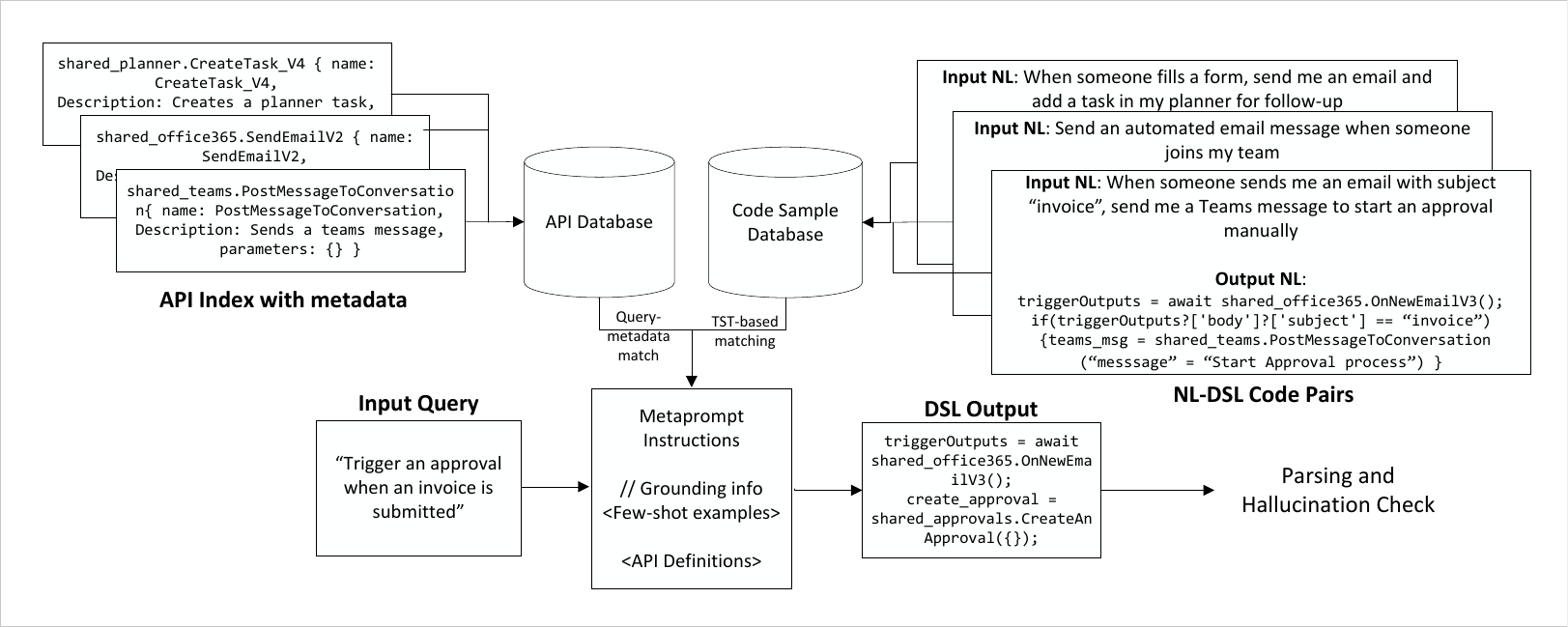}
  \caption{System Architecture to show e2e working of
    our DSL generation methodology using RAG. TST based semantic mapping
    retrieves the relevant code snippet as shown. This helps get the right syntax. However, it gets the correct function name for approval from the API metadata}
  \label{fig:system_arch}
\end{figure*}

Domain Specific Languages (or DSLs) are custom computer languages designed and optimized for specific applications. Examples of DSLs include SQL and industry-specific languages for formalizing API calls, often using formats like JSON or YAML to represent API sequences. In this paper, we focus on the task of generating a DSL used for authoring high-level automation workflows across thousands of web-scale APIs. These workflows support a variety of customer scenarios like invoice processing, sales lead integration with forms/emails etc. The automation DSL represents API names as functions and codifies a sequence of API calls (or a plan) along with conditional logic over the invocation of APIs. The syntax itself borrows from known languages like Javascript, however, the logic resembling the workflow along with the custom function names, make it unique. An example of the DSL is shown in Figure \ref{fig:system_arch}.

Existing code generation methods are hard to adapt for this scenario due to the frequent hallucinations and syntax errors. This is largely due to the custom API names, high cardinality and diversity of APIs in public as well private domain along with the ever-changing API landscape. A typical workflow can choose among thousands of publicly available APIs (eg. Database connectors, Emails, Planner, Notifications etc.) as well as private APIs in tenant subscriptions and string them together to automate business processes. 

In this paper, we outline an end to end system architecture for NL2DSL generation with high response rate using selective improvements to RAG techniques (\citealp{Liu23ChatGPTPrompts}, \citealp{Poesia22TST}) using OpenAI models. We focus on bench-marking the impact of different contexts used for grounding. We fine-tuned a Codex model for NL2DSL and show a comparative analysis of the impact of the approaches used to optimize RAG. 

The remainder of this study is structured as follows. In Section \ref{relatedwork}, we present the NL2DSL problem formulation along with literature review. Section \ref{methodology} lays out and describes the optimizations we made to RAG as discussed above along with the benchmark Fine-Tuned model. Section \ref{expresults} discusses Data Generation, Metric definition and Section \ref{results} shares our results and discussion followed by Conclusion in Section \ref{conclusion}.

\section{Related Work} \label{relatedwork}

\subsection{Code Generation or Program Synthesis}

Program Synthesis is a hard research problem (\citealp{jain22jigsaw}, \citealp{Li_2022}, \citealp{xu2021inide}). It has gained significant interest with many open-source models (\citealp{CodeLlama23}, \citealp{li2023starcoder}, \citealp{Codestral24}, \citealp{abdin2024phi3}, \citealp{Chen21CodeGenEval}) focusing on general programming languages. Many of these advancements have been achieved through pre-training language models for code generation with a focus on improving datasets (\citealp{CodeLlama23}, \citealp{abdin2024phi3}). However, for domain adaptation, \textbf{instruction fine-tuning} on top of a base model remains a popular approach (\citealp{Chen21CodeGenEval}, \citealp{gao2023pal}, \citealp{lewkowycz2022solving}, \citealp{Patil23Gorilla}). 

Prompting LLMs is an alternative technique for code generation (\citealp{Liu23ChatGPTPrompts}, \citealp{White23ChatGPTPrompts}, \citealp{wei2023chainofthought}, \citealp{kojima2023large}). Most papers focus on metaprompt optimization and learning, while \citealp{Poesia22TST} focused on improving response quality by dynamically selecting few-shots for grounding the model.

\subsection{Reasoning and Tool Integration}
For modeling the problem of selecting a sequence of API calls, we need to consider formulating it as a \textbf{planning} or \textbf{reasoning} task. LLMs show remarkable reasoning capability, however, they also have limitations when it comes to staying up-to-date with recent knowledge, performing mathematical calculations etc. A popular way to overcome this has been granting the LLMs access to external tools. This framework gained significant popularity with OpenAI Code Interpreter's success (\citealp{CodeInter23}). 

External Tool Integration has been studied since with a focus on including specific tools such as web search (\citealp{Schick23Toolformer}), python code interpreters (\citealp{gao2023pal}, \citealp{CodeInter23}), adding calculators (\citealp{Parisi2022Talm} \citealp{gao2023pal}) and so on. Expanding the tool set to a generic list of tools has been explored (\citealp{Schick23Toolformer}, \citealp{Patil23Gorilla}), but it remains limited and often predicts single tool instead of sequences needed for most enterprise scenarios. Tool Use has mostly been explored in the context of generating more accurate text outputs for Q\&A tasks with the help of external tools(\citealp{Schick23Toolformer}, \citealp{Parisi2022Talm}).

There is an increase in focus on incorporating LLM's code generation capabilities to reasoning and task orchestration making this an area of active research (\citealp{gao2023pal}, \citealp{liang2023taskmatrixai}, \citealp{Patil23Gorilla}). However, similar to Q\&A scenarios mentioned above, most of the research either limits the tools to a set of small well-documented APIs (\citealp{gao2023pal}, \citealp{liang2023taskmatrixai}), or limited their scope to predicting a single output API (\citealp{Patil23Gorilla}, \citealp{Schick23Toolformer}). 

Posing the reasoning or orchestration task as a code generation problem is similar to the API sequence generation scenario highlighted in this paper. Improving the quality of Natural Language to DSL generation, is thus beneficial for both reasoning and plan generation.

\subsection{Contributions}
NL2DSL generation suffers from the hallucination and quality issues we discussed in Section \ref{intro}. Few studies address the challenges of end-to-end DSL generation, specifically over a large set of custom APIs. In this paper, we present an end-to-end system architecture with improved strategies to add grounding context with known RAG techniques. We also present an ablation study showing improvements for DSL generation quality for enterprise settings. DSL samples in our test set consider API or tool selection sequences of 5-6 API calls, also referred to as chain of tools, which is a first to the best of our knowledge. We also consider the real-world scenarios of adding conditional logic with API calls as shown with an example in Figure \ref{fig:system_arch}. 

\section{Methodology} \label{methodology}
\subsection{Fine-Tuned NL2DSL Generation Model} \label{ss:fine-tuning}
We took the Codex base model from OpenAI due to it's pre-training with code samples and used LoRA-based fine-tuning approach. The training set consists of NL-DSL pairs, NL refers to the user query and the DSL represents the workflow that the user is looking to automate. The training set consists of a pool of 67k samples in the form of (prompt, flow) tuples with the NL generated synthetically (details in Section \ref{ss:dataset}, examples of NL-DSL are shared in Figure \ref{fig:system_arch} and Appendix \ref{sec:appendix}).

We ran many iterations on this model to improve performance on the test set, specifically for the body and tail connectors, and went through multiple rounds of data augmentation. We found that predicting the parameter keys was very challenging with the fine-tuned model due to limitation of high-quality data generation. 

\subsection{Grounding with dynamically selected few-shots} \label{ss:RAG}
We tried two types of grounding information for RAG based DSL generation as described below. For each technique, we selected $5$ and $20$ few-shots dynamically, and compared performance impact driven by the approach used for sample selection.

\subsubsection{Pre-trained Model}
The first approach is using a vanilla Pre-trained model for determining the semantic similarity of NL-DSL samples based on the NL query. We computed the embeddings of NL queries using a Distil-RoBERTa Pre-trained model. We created a Faiss Index (\citealp{Faiss2024}) for these embeddings to help with search over the dense embedding space.

\subsubsection{TST based BERT Fine-tuning}
In this approach, we fine-tuned the Pre-trained model to improve retrieval accuracy of few-shots similar to the work done by \citealp{Poesia22TST}.

To get positive and negative samples for fine-tuning, we generated embeddings for all NL queries in our dataset using a Pre-trained Tansformer model. A pair of tuples is considered a positive sample if the cosine similarity between their corresponding NL prompts is greater than $0.7$ and negative otherwise. We generated $100k$ pairs this way and leveraged them as training data for our fine-tuned model. 

The loss function used by TST (Equation \ref{MSE_loss} from \citealp{Poesia22TST}) is minimizing the Mean-Squared Error between the vanilla loss functions comparing the utterances ($u_i,u_j$) and the target programs ($p_i,p_j$). Program similarity is denoted by $S$. We used a Jaccard score over lists of API function names as the similarity metric between programs. 
\begin{equation} \label{MSE_loss}
L_{TST}(\theta):= E_{i,j ~ D} [f_{\theta}(u_i, u_j) - S (P_i,p_j)]^2
\end{equation}

\subsection{Grounding with API Metadata} \label{ss:tool_defs}
In addition to few-shots, we appended the API metadata in the metaprompt. This metadata includes Function Description along with the parameter keys and their description (Example API Function Definition shared in Appendix \ref{sec:appendix}). We followed the below two approaches for selecting the metadata to be added. 

\subsubsection{API Function Definitions for Few-Shots}
For few-shot samples selected using the methods described above, we extracted the metadata for \textbf{each of} the functions present in those samples. This means that for the $n$ few-shot samples dynamically added to the metaprompt, we iterated over all the API function names in each of these flows and added their function definitions to the metaprompt. 

\subsubsection{Semantic Function Definitions}
Another approach for selecting function definitions is to retrieve semantically similar functions from a vector database created with API metadata. This approach is similar to the one followed by (\citealp{LlamaIndex}). We created an index of all API definitions and used the input NL query for search. Please note that this is different from the faiss index created for few-shot samples in Section \ref{ss:RAG}.

We call this approach \textbf{Semantic Function Definition (SFD)} and will compare it with the \textbf{Regular FDs} described above. This approach can be specifically useful for tail-ish prompts where no few-shots might be retrieved.

\section{Experiment Design and Metrics Definition} \label{expresults}
In this section, we outline the process of Dataset Generation and introduce the metrics we used for estimating the code quality. We then describe our experiments. We have used Azure AML pipelines and GPT-4 (16k token limit) for our experiments.

\begin{table*}
  \begin{tabular}{lccccc}
  \hline
    Model & \parbox{2 cm}{Num. of \\few-shots} & \parbox{2.5 cm}{Avg. similarity} & \parbox{2 cm}{\%Unparsed flows} & \parbox{2 cm}{\%Made-up API names} & \parbox{2 cm}{\%Made-up parameters}\\
    \hline
    \textbf{{Pre-trained w\/o FD}} & 20&\boldmath{$+0.03$}& \boldmath{$-3.37$}& $-7.34$& \boldmath{$-15.17$}\\
    \textbf{TST w\/o FD}& 5&$+0.02$& $-0.61$& $-3.53$& $-1.04$\\
    \textbf{TST w\/o FD}&20&\boldmath{$+0.03$}& $-2.85$& \boldmath{$-8.49$}& $-14.58$\\
    \hline
  \end{tabular}
  \caption{Impact of selecting \textbf{5 vs 20 few-shot} samples for TST and Pre-trained Model without adding API Function Definitions using GPT-4. The baseline uses Pre-trained Transformer Model with 5 few-shot samples. }
  \label{tab:fewshotcomparisonmodel5n20}
\end{table*}

\subsection{Dataset Generation} \label{ss:dataset}
Our train and test set consists of a total of 67k and 1k samples, respectively. These samples are (prompt, flow) pairs with the workflows being created by users across a large set of APIs. We scrubbed PII from these automations and sampled workflows containing $700$ publicly available APIs. We synthetically generated the corresponding Natural Language prompts using GPT-4.

\subsection{DSL Generation Quality Metrics}
We defined 3 key metrics to focus on code generation quality as well as syntactic accuracy and hallucination rate.

\subsubsection{Average Similarity} 
Average Similarity measures the similarity between predicted flow and the ground truth flow and is defined using the Longest Common Subsequence match (LCSS) metric. Each flow is reduced to a list of API sequences and then the LCSS is computed. The final metric is reported as an average over all test samples. Hallucination and Parser failures lead to the sample being discarded and is assigned a similarity score of 0.

\begin{equation}
    \textrm{Similarity} = \frac{\mathrm{LCSS} (A, B)} {max (|\mathrm{Actions}_A|, |\mathrm{Actions}_B|)} 
\end{equation}
where $|\textrm{Actions}_A|$ is the number of actions in flow $A$ and $|\textrm{Actions}_B|$ is the number of actions in flow $B$.
 
\begin{table*}
  \begin{tabular}{lcccc}
   \hline
    Model & Avg. Similarity & \%Unparsed flows & \%Made-up API names & \parbox{2.5cm}{\%Made-up API parameters}\\
    \hline
    \textbf{Pre-trained + FD} & $0$ & $+2.75$ & $-4.3$ & \boldmath{$-20.16$} \\
    \textbf{TST w\/o FD} & \boldmath{$+0.02$} & \boldmath{$-0.61$} & $-3.53$ & $-1.04$ \\
    \textbf{TST + FD} & \boldmath{$+0.02$} & $+0.68$ & \boldmath{$-6.29$} & $-19.99$ \\
    \hline
  \end{tabular}
  \caption{Impact of selecting \textbf{5 few-shot} samples using TST vs. Pre-trained Model with and without API Function Definitions using GPT4 model. The baseline uses Pre-trained Transformer Model without API Function Definitions.} 
  \label{tab:fewshotselectionmodel}
\end{table*}

\subsubsection{Unparsed rate}
This metric captures the rate of syntactic errors. A flow that cannot be parsed by the parser is considered not usable for the purpose of similarity metric computation. Unparsed rate is computed as follow:

\begin{equation}
    \%\mathrm{unparsed \ flows} = \frac{|\mathrm{Flows}_\mathrm{unparsed}|}{|\mathrm{Flows}_\mathrm{total}|}
\end{equation}
where, $|\mathrm{Flows}_\mathrm{unparsed}|$ is the number of flows that were not parsed and $|\mathrm{Flows}_\mathrm{total}|$is the total number of flows in the sample set.
        
\subsubsection{Hallucination rate}
This metric captures the rate of made-up APIs and made-up parameter keys in the generated code. Predicting a flow with a hallucinated API name is counted as a failure and leads to the code being considered invalid. However, hallucinated parameter keys would only lead to run-time errors which can be fixed down the line. Fixing these run-time errors is beyond the scope of this paper.

\begin{equation}
    \%\mathrm{made-up \ APIs} = \frac{|\mathrm{Flows}_h|}{|\mathrm{Flows}_\mathrm{parsed}|} * 100
\end{equation}

\begin{equation}
    \%\mathrm{made-up \ parameters} = \frac{|\mathrm{Flows}_{hp}|}{|\mathrm{Flows}_\mathrm{parsed}|} * 100
\end{equation}
where $|\mathrm{Flows}_h|$ is the number of flows with hallucinated API names, $|\mathrm{Flows}_{hp}|$ is the number of flows with hallucinated parameter key names and $|\mathrm{Flows}_\mathrm{parsed}|$ is the number of flows that were parsed correctly.

\section{Results} \label{results}
In this section, we present the results of the above approaches on a test set of 1000 NL-DSL pairs. The test set is split in $864$ in-domain samples and $136$ out-of-domain samples. The NL component in these samples, while generated synthetically, has been evaluated by human judges for quality. They were also sampled to represent the distribution of APIs in actual product usage.

We compare the impact of each ablation in sections below. Please note that in the following sections all the results are presented as $\Delta$ change compared to a baseline scenario where the higher $\Delta$ is better for Avg. similarity and lower $\Delta$ is better for the rest of metrics capturing failures.

\subsection{Impact of number of few-shots}
We compare the impact of number of code samples added to the meta prompt with two different settings i.e. $5$ few-shots vs $20$ few-shots. We measured the results for both Pre-trained model as well as TST model. Results are shared in Table \ref{tab:fewshotcomparisonmodel5n20}.

Looking at rows 2 and 4 having 20 few-shot samples, we can see that adding more few-shots improves the performance of both the Pre-trained as well as the TST model on all metrics. The gain is particularly pronounced for reducing the number of made-up API names as well as reducing the number of made-up API parameter keys.
\begin{table*}
  \begin{tabular}{lcccc}
   \hline
    Model & \parbox{2.5cm}{Avg. Similarity} & \parbox{3cm}{\%Unparsed flows} & \parbox{2.75cm}{\%Made-up API names} & \parbox{2.75cm}{\%Made-up API parameters}\\
    \hline
    \textbf{Pre-trained + FD} & $-0.01$ & $+2.29$ & $-2.17$ & $-6.93$\\
    \textbf{TST w\/o FD} & $0$ & \boldmath{$+0.52$} & $-1.15$ & $+0.52$\\
    \textbf{TST + FD} & \boldmath{$+0.02$} & $+0.83$ & \boldmath{$-2.7$} & \boldmath{$-7.06$}\\
    \hline
  \end{tabular}
  \caption{Impact of selecting \textbf{20 few-shot} samples using TST vs. Pre-trained Model with and without Function Definitions using GPT4 model. The baseline uses Pre-trained Transformer Model without API Function Definitions.}
  \label{tab:fewshotselectionmodel20}
\end{table*}

\begin{table*}
  \begin{tabular}{lcccc}
   \hline
    Model & Avg. Similarity & \%Unparsed flows& \%Made-up API names& \parbox{2.5cm}{\%Made-up API parameters}\\
    \hline
    \textbf{TST + FD} & \boldmath{$0$}& \boldmath{$-5.3$}& $+1.7$& \boldmath{$+1.11$}\\
    \textbf{TST + SFD} & $-0.01$& $-1.43$& $+1.21$& $+6.76$\\
    \textbf{TST + FD + SFD} &\boldmath{$0$} & $-2.74$& \boldmath{$+0.94$}& $+2.03$\\
    \hline
  \end{tabular}
  \caption{Impact of adding API or tool related metadata on performance (with GPT-4 model and 20 few-shots). FD refers to including only metadata for APIs present in few-shots. SFD refers to extracting APIs similar to the input query (refer to Section \ref{methodology}) for details. The baseline uses fine-tuned Codex model.}
  \label{tab:apiselectionmodel}
\end{table*}

\begin{table*}
  \begin{tabular}{lcccc}
   \hline
    Test set & Avg. Similarity & \%Unparsed flows & \%Made-up API names & \parbox{3.15cm}{\%Made-up API \\ parameters}\\
    \hline
    \textbf{OOD test set} & \boldmath{$+0.07$} & $+12.23$ & \boldmath{$-1.98$} & $+13.11$\\
    \textbf{Full test set} & $0$ & $+1.06$ & \boldmath{$-0.48$} & $+3.88$\\
    \hline
  \end{tabular}
   \caption{Performance of RAG based model (TST + FD + SFD) on out of domain (OOD) and full test sets (with GPT-4 model and 20 few-shots). The baseline is fine-tuned Codex model on updated training data.}
  \label{tab:tailset}
\end{table*}

\subsection{TST vs Pre-trained Model}
To compare the impact of selecting samples using TST vs Pre-trained model, we look at the impact with and without the presence of API Function Definitions (see Table \ref{tab:fewshotselectionmodel} and Table \ref{tab:fewshotselectionmodel20}). Here, we have used GPT4 model with 5 and 20 few-shots, respectively. TST with FD setting performs overall better than all other options with values close to the best in every metric.

This leads us to conclude that the presence of few-shot examples is supported by adding their API functions definitions (as described in Section \ref{methodology}). The addition predominantly helps reducing the hallucination rate for API names and parameters, which improves the overall response rate of NL2DSL generation. This supports our initial premise: adding tool descriptions (like it is done in planning tasks) along with few-shot code samples helps improve reliability of plan generation. 

\subsection{Regular Function Definition vs Semantic Function Definitions} 
We used a Fine-Tuned model as baseline for this experiment (Table \ref{tab:apiselectionmodel}). Based on the insights from the previous step, we used 20 few-shots for TST along with including FDs. 

Looking at metrics in columns for $\%$ made-up API names and $\%$ made-up parameter keys, we see that the hallucination rate is, in general, increasing for RAG based approach. However, we need to keep in mind that a fine-tuned model on the function names is hard to beat as it has been trained on 67,000 samples compared to only 20 few-shots that have been added to the RAG model.

Within the RAG approaches, comparing rows 1 and 2 ("TST + FD" vs "TST + SFD"), SFD results in a slight drop in average similarity and an increase in the unparsed rate and hallucination rate for parameters. This indicates that the approach to simply add semantically similar API metadata for a query is not useful for DSL generation. We get better similarity, and reduced hallucination rate, when we include the API Function Definitions for the samples selected by TST (as shown in Row 1).

\subsection{Out of Domain APIs} 
To compare the impact of RAG on unseen APIs, not available for fine-tuning, we created an out of domain test set. We selected 10 APIs, and discarded the flows containing these APIs from the train set. The test set contains 136 (NL, flow) pairs having these APIs.

We share the results in Table \ref{tab:tailset}. The baseline is a fine-tuned Codex model with the updated training data. The RAG-based approach notably enhances average similarity (by 7 pts) and reduces API hallucinations (by 1.5 pts) for out of domain APIs. This indicates that when samples are not present in the train set, grounding with RAG context can provide the LLM support for improving code quality.

However, fine-tuned model outperforms RAG model in terms of syntactic errors and parameter key hallucinations. The role of few-shots in informing the syntax of the output code cannot be substituted with just adding function definitions. Since, it is hard to obtain the examples for unseen APIs, we need to find alternate ways to improve syntactic errors. We will look into improving this as future work.

\section{Conclusion} \label{conclusion}
Based on the presented results, we see that the role of dynamically selected few-shot samples is very important in making RAG useful for syntactically correct generation of DSL as well as improving code similarity (Table \ref{tab:apiselectionmodel}). This could be due to the fact that few-shot examples have been successfully teaching the correct syntax to the LLM model. The positive role of relevant few-shot samples in improving RAG's syntactic accuracy is further confirmed by the drop seen for out of domain data. In absence of relevant few-shots for unseen APIs, we chose examples with low similarity, directly impacting the syntactic accuracy (Table \ref{tab:tailset}). 

Counter intuitively, with the exception of out-of-domain data, this benefit does not transfer to hallucinated API names and their parameter keys where the fine-tuned model holds the advantage (Table \ref{tab:apiselectionmodel}). Among RAG approaches (Tables \ref {tab:fewshotselectionmodel} and \ref{tab:fewshotselectionmodel20}), TST + Regular Function Definitions reduced hallucinations the most. Adding Semantic Function Definitions to TST + FD did not confer any advantage for in-domain APIs, but greatly improved code similarity for out-of domain APIs. 

Overall, we were able to significantly improve the performance of RAG for DSL generation, with hallucination rate for API names dropping by 6.29 pts. and by 20 pts for parameter keys (Table \ref{tab:fewshotselectionmodel}). The performance of RAG is now comparable to that of fine-tuned model (see Avg. Similarity in Table \ref{tab:apiselectionmodel}), with better performance for unseen APIs. This reduces the need to fine-tune the model frequently for new APIs saving compute and resources. 




\section {Ethical Considerations}

We used instructions in meta prompt to not respond to harmful queries. This is supplemented with a harms classifier on the input prompt. The Fine-tuned model was shown examples where it should not generate an output and consequently learnt not to respond to queries considered harmful.

\bibliography{custom}

\begin{thebibliography}{25}
\providecommand{\natexlab}[1]{#1}

\bibitem[{Abdin et~al.(2024)Abdin, Jacobs, Awan, Aneja, Awadallah, Awadalla, Bach, Bahree, Bakhtiari, Bao, Behl, Benhaim, Bilenko, Bjorck, Bubeck, Cai, Cai, Mendes, Chen, Chaudhary, Chen, Chen, Chen, Chen, Chopra, Dai, Giorno, de~Rosa, Dixon, Eldan, Fragoso, Iter, Gao, Gao, Gao, Garg, Goswami, Gunasekar, Haider, Hao, Hewett, Huynh, Javaheripi, Jin, Kauffmann, Karampatziakis, Kim, Khademi, Kurilenko, Lee, Lee, Li, Li, Liang, Liden, Liu, Liu, Liu, Lin, Lin, Luo, Madan, Mazzola, Mitra, Modi, Nguyen, Norick, Patra, Perez-Becker, Portet, Pryzant, Qin, Radmilac, Rosset, Roy, Ruwase, Saarikivi, Saied, Salim, Santacroce, Shah, Shang, Sharma, Shukla, Song, Tanaka, Tupini, Wang, Wang, Wang, Wang, Ward, Wang, Witte, Wu, Wyatt, Xiao, Xu, Xu, Xu, Yadav, Yang, Yang, Yang, Yang, Yu, Yuan, Zhang, Zhang, Zhang, Zhang, Zhang, Zhang, Zhang, and Zhou}]{abdin2024phi3}
Marah Abdin, Sam~Ade Jacobs, Ammar~Ahmad Awan, Jyoti Aneja, Ahmed Awadallah, Hany Awadalla, Nguyen Bach, Amit Bahree, Arash Bakhtiari, Jianmin Bao, Harkirat Behl, Alon Benhaim, Misha Bilenko, Johan Bjorck, Sébastien Bubeck, Qin Cai, Martin Cai, Caio César~Teodoro Mendes, Weizhu Chen, Vishrav Chaudhary, Dong Chen, Dongdong Chen, Yen-Chun Chen, Yi-Ling Chen, Parul Chopra, Xiyang Dai, Allie~Del Giorno, Gustavo de~Rosa, Matthew Dixon, Ronen Eldan, Victor Fragoso, Dan Iter, Mei Gao, Min Gao, Jianfeng Gao, Amit Garg, Abhishek Goswami, Suriya Gunasekar, Emman Haider, Junheng Hao, Russell~J. Hewett, Jamie Huynh, Mojan Javaheripi, Xin Jin, Piero Kauffmann, Nikos Karampatziakis, Dongwoo Kim, Mahoud Khademi, Lev Kurilenko, James~R. Lee, Yin~Tat Lee, Yuanzhi Li, Yunsheng Li, Chen Liang, Lars Liden, Ce~Liu, Mengchen Liu, Weishung Liu, Eric Lin, Zeqi Lin, Chong Luo, Piyush Madan, Matt Mazzola, Arindam Mitra, Hardik Modi, Anh Nguyen, Brandon Norick, Barun Patra, Daniel Perez-Becker, Thomas Portet, Reid Pryzant, Heyang
  Qin, Marko Radmilac, Corby Rosset, Sambudha Roy, Olatunji Ruwase, Olli Saarikivi, Amin Saied, Adil Salim, Michael Santacroce, Shital Shah, Ning Shang, Hiteshi Sharma, Swadheen Shukla, Xia Song, Masahiro Tanaka, Andrea Tupini, Xin Wang, Lijuan Wang, Chunyu Wang, Yu~Wang, Rachel Ward, Guanhua Wang, Philipp Witte, Haiping Wu, Michael Wyatt, Bin Xiao, Can Xu, Jiahang Xu, Weijian Xu, Sonali Yadav, Fan Yang, Jianwei Yang, Ziyi Yang, Yifan Yang, Donghan Yu, Lu~Yuan, Chengruidong Zhang, Cyril Zhang, Jianwen Zhang, Li~Lyna Zhang, Yi~Zhang, Yue Zhang, Yunan Zhang, and Xiren Zhou. 2024.
\newblock \href {https://arxiv.org/abs/2404.14219} {Phi-3 technical report: A highly capable language model locally on your phone}.
\newblock \emph{Preprint}, arXiv:2404.14219.

\bibitem[{ChatGPT()}]{ChatGPT22}
ChatGPT. 2022.
\newblock \href {https://openai.com/blog/chatgpt} {Chatgpt}.

\bibitem[{Chen et~al.(2021)Chen, Tworek, Jun, Yuan, de~Oliveira~Pinto, Kaplan, Edwards, Burda, Joseph, Brockman, Ray, Puri, Krueger, Petrov, Khlaaf, Sastry, Mishkin, Chan, Gray, Ryder, Pavlov, Power, Kaiser, Bavarian, Winter, Tillet, Such, Cummings, Plappert, Chantzis, Barnes, Herbert-Voss, Guss, Nichol, Paino, Tezak, Tang, Babuschkin, Balaji, Jain, Saunders, Hesse, Carr, Leike, Achiam, Misra, Morikawa, Radford, Knight, Brundage, Murati, Mayer, Welinder, McGrew, Amodei, McCandlish, Sutskever, and Zaremba}]{Chen21CodeGenEval}
Mark Chen, Jerry Tworek, Heewoo Jun, Qiming Yuan, Henrique~Ponde de~Oliveira~Pinto, Jared Kaplan, Harri Edwards, Yuri Burda, Nicholas Joseph, Greg Brockman, Alex Ray, Raul Puri, Gretchen Krueger, Michael Petrov, Heidy Khlaaf, Girish Sastry, Pamela Mishkin, Brooke Chan, Scott Gray, Nick Ryder, Mikhail Pavlov, Alethea Power, Lukasz Kaiser, Mohammad Bavarian, Clemens Winter, Philippe Tillet, Felipe~Petroski Such, Dave Cummings, Matthias Plappert, Fotios Chantzis, Elizabeth Barnes, Ariel Herbert-Voss, William~Hebgen Guss, Alex Nichol, Alex Paino, Nikolas Tezak, Jie Tang, Igor Babuschkin, Suchir Balaji, Shantanu Jain, William Saunders, Christopher Hesse, Andrew~N. Carr, Jan Leike, Josh Achiam, Vedant Misra, Evan Morikawa, Alec Radford, Matthew Knight, Miles Brundage, Mira Murati, Katie Mayer, Peter Welinder, Bob McGrew, Dario Amodei, Sam McCandlish, Ilya Sutskever, and Wojciech Zaremba. 2021.
\newblock \href {https://arxiv.org/abs/2107.03374} {Evaluating large language models trained on code}.
\newblock \emph{Preprint}, arXiv:2107.03374.

\bibitem[{Code Llama()}]{CodeLlama23}
Code Llama. 2023.
\newblock \href {https://doi.org/10.48550/arXiv.2308.12950} {Code llama: Open foundation models for code}.

\bibitem[{Codestral()}]{Codestral24}
Codestral. 2024.
\newblock \href {https://mistral.ai/news/codestral/} {Codestral}.

\bibitem[{Douze et~al.(2024)Douze, Guzhva, Deng, Johnson, Szilvasy, Mazaré, Lomeli, Hosseini, and Jégou}]{Faiss2024}
Matthijs Douze, Alexandr Guzhva, Chengqi Deng, Jeff Johnson, Gergely Szilvasy, Pierre-Emmanuel Mazaré, Maria Lomeli, Lucas Hosseini, and Hervé Jégou. 2024.
\newblock \href {https://arxiv.org/abs/2401.08281} {The faiss library}.
\newblock \emph{Preprint}, arXiv:2401.08281.

\bibitem[{Gao et~al.(2023)Gao, Madaan, Zhou, Alon, Liu, Yang, Callan, and Neubig}]{gao2023pal}
Luyu Gao, Aman Madaan, Shuyan Zhou, Uri Alon, Pengfei Liu, Yiming Yang, Jamie Callan, and Graham Neubig. 2023.
\newblock \href {https://arxiv.org/abs/2211.10435} {Pal: Program-aided language models}.
\newblock \emph{Preprint}, arXiv:2211.10435.

\bibitem[{GPT-4()}]{GPT423}
GPT-4. 2023.
\newblock \href {https://cdn.openai.com/papers/gpt-4.pdf} {Gpt-4 technical report}.

\bibitem[{Jain et~al.(2021)Jain, Vaidyanath, Iyer, Natarajan, Parthasarathy, Rajamani, and Sharma}]{jain22jigsaw}
Naman Jain, Skanda Vaidyanath, Arun Iyer, Nagarajan Natarajan, Suresh Parthasarathy, Sriram Rajamani, and Rahul Sharma. 2021.
\newblock \href {https://arxiv.org/abs/2112.02969} {Jigsaw: Large language models meet program synthesis}.
\newblock \emph{Preprint}, arXiv:2112.02969.

\bibitem[{Kojima et~al.(2023)Kojima, Gu, Reid, Matsuo, and Iwasawa}]{kojima2023large}
Takeshi Kojima, Shixiang~Shane Gu, Machel Reid, Yutaka Matsuo, and Yusuke Iwasawa. 2023.
\newblock \href {https://arxiv.org/abs/2205.11916} {Large language models are zero-shot reasoners}.
\newblock \emph{Preprint}, arXiv:2205.11916.

\bibitem[{Lewkowycz et~al.(2022)Lewkowycz, Andreassen, Dohan, Dyer, Michalewski, Ramasesh, Slone, Anil, Schlag, Gutman-Solo, Wu, Neyshabur, Gur-Ari, and Misra}]{lewkowycz2022solving}
Aitor Lewkowycz, Anders Andreassen, David Dohan, Ethan Dyer, Henryk Michalewski, Vinay Ramasesh, Ambrose Slone, Cem Anil, Imanol Schlag, Theo Gutman-Solo, Yuhuai Wu, Behnam Neyshabur, Guy Gur-Ari, and Vedant Misra. 2022.
\newblock \href {https://arxiv.org/abs/2206.14858} {Solving quantitative reasoning problems with language models}.
\newblock \emph{Preprint}, arXiv:2206.14858.

\bibitem[{Li et~al.(2023)Li, Allal, Zi, Muennighoff, Kocetkov, Mou, Marone, Akiki, Li, Chim, Liu, Zheltonozhskii, Zhuo, Wang, Dehaene, Davaadorj, Lamy-Poirier, Monteiro, Shliazhko, Gontier, Meade, Zebaze, Yee, Umapathi, Zhu, Lipkin, Oblokulov, Wang, Murthy, Stillerman, Patel, Abulkhanov, Zocca, Dey, Zhang, Fahmy, Bhattacharyya, Yu, Singh, Luccioni, Villegas, Kunakov, Zhdanov, Romero, Lee, Timor, Ding, Schlesinger, Schoelkopf, Ebert, Dao, Mishra, Gu, Robinson, Anderson, Dolan-Gavitt, Contractor, Reddy, Fried, Bahdanau, Jernite, Ferrandis, Hughes, Wolf, Guha, von Werra, and de~Vries}]{li2023starcoder}
Raymond Li, Loubna~Ben Allal, Yangtian Zi, Niklas Muennighoff, Denis Kocetkov, Chenghao Mou, Marc Marone, Christopher Akiki, Jia Li, Jenny Chim, Qian Liu, Evgenii Zheltonozhskii, Terry~Yue Zhuo, Thomas Wang, Olivier Dehaene, Mishig Davaadorj, Joel Lamy-Poirier, João Monteiro, Oleh Shliazhko, Nicolas Gontier, Nicholas Meade, Armel Zebaze, Ming-Ho Yee, Logesh~Kumar Umapathi, Jian Zhu, Benjamin Lipkin, Muhtasham Oblokulov, Zhiruo Wang, Rudra Murthy, Jason Stillerman, Siva~Sankalp Patel, Dmitry Abulkhanov, Marco Zocca, Manan Dey, Zhihan Zhang, Nour Fahmy, Urvashi Bhattacharyya, Wenhao Yu, Swayam Singh, Sasha Luccioni, Paulo Villegas, Maxim Kunakov, Fedor Zhdanov, Manuel Romero, Tony Lee, Nadav Timor, Jennifer Ding, Claire Schlesinger, Hailey Schoelkopf, Jan Ebert, Tri Dao, Mayank Mishra, Alex Gu, Jennifer Robinson, Carolyn~Jane Anderson, Brendan Dolan-Gavitt, Danish Contractor, Siva Reddy, Daniel Fried, Dzmitry Bahdanau, Yacine Jernite, Carlos~Muñoz Ferrandis, Sean Hughes, Thomas Wolf, Arjun Guha, Leandro von
  Werra, and Harm de~Vries. 2023.
\newblock \href {https://arxiv.org/abs/2305.06161} {Starcoder: may the source be with you!}
\newblock \emph{Preprint}, arXiv:2305.06161.

\bibitem[{Li et~al.(2022)Li, Choi, Chung, Kushman, Schrittwieser, Leblond, Eccles, Keeling, Gimeno, Dal~Lago, Hubert, Choy, de~Masson~d’Autume, Babuschkin, Chen, Huang, Welbl, Gowal, Cherepanov, Molloy, Mankowitz, Sutherland~Robson, Kohli, de~Freitas, Kavukcuoglu, and Vinyals}]{Li_2022}
Yujia Li, David Choi, Junyoung Chung, Nate Kushman, Julian Schrittwieser, Rémi Leblond, Tom Eccles, James Keeling, Felix Gimeno, Agustin Dal~Lago, Thomas Hubert, Peter Choy, Cyprien de~Masson~d’Autume, Igor Babuschkin, Xinyun Chen, Po-Sen Huang, Johannes Welbl, Sven Gowal, Alexey Cherepanov, James Molloy, Daniel~J. Mankowitz, Esme Sutherland~Robson, Pushmeet Kohli, Nando de~Freitas, Koray Kavukcuoglu, and Oriol Vinyals. 2022.
\newblock \href {https://doi.org/10.1126/science.abq1158} {Competition-level code generation with alphacode}.
\newblock \emph{Science}, 378(6624):1092–1097.

\bibitem[{Liang et~al.(2023)Liang, Wu, Song, Wu, Xia, Liu, Ou, Lu, Ji, Mao, Wang, Shou, Gong, and Duan}]{liang2023taskmatrixai}
Yaobo Liang, Chenfei Wu, Ting Song, Wenshan Wu, Yan Xia, Yu~Liu, Yang Ou, Shuai Lu, Lei Ji, Shaoguang Mao, Yun Wang, Linjun Shou, Ming Gong, and Nan Duan. 2023.
\newblock \href {https://arxiv.org/abs/2303.16434} {Taskmatrix.ai: Completing tasks by connecting foundation models with millions of apis}.
\newblock \emph{Preprint}, arXiv:2303.16434.

\bibitem[{Liu et~al.(2023)Liu, Bao, Zhang, Zhang, Hu, Zhang, and Yan}]{Liu23ChatGPTPrompts}
Chao Liu, Xuanlin Bao, Hongyu Zhang, Neng Zhang, Haibo Hu, Xiaohong Zhang, and Meng Yan. 2023.
\newblock \href {https://arxiv.org/abs/2305.08360} {Improving chatgpt prompt for code generation}.
\newblock \emph{Preprint}, arXiv:2305.08360.

\bibitem[{LlamaIndex()}]{LlamaIndex}
LlamaIndex. 2023.
\newblock \href {https://llama.meta.com/docs/integration-guides/llamaindex/} {Llamaindex}.

\bibitem[{Nguyen and Nadi(2022)}]{Nguyen22Eval}
Nhan Nguyen and Sarah Nadi. 2022.
\newblock \href {https://doi.org/10.1145/3524842.3528470} {An empirical evaluation of github copilot's code suggestions}.
\newblock In \emph{2022 IEEE/ACM 19th International Conference on Mining Software Repositories (MSR)}, pages 1--5.

\bibitem[{OpenAI Code Interpretor()}]{CodeInter23}
OpenAI Code Interpretor. 2023.
\newblock \href {https://platform.openai.com/docs/assistants/tools/code-interpreter} {Openai code interpretor}.

\bibitem[{Parisi et~al.(2022)Parisi, Zhao, and Fiedel}]{Parisi2022Talm}
Aaron Parisi, Yao Zhao, and Noah Fiedel. 2022.
\newblock \href {https://arxiv.org/abs/2205.12255} {Talm: Tool augmented language models}.
\newblock \emph{Preprint}, arXiv:2205.12255.

\bibitem[{Patil et~al.(2023)Patil, Zhang, Wang, and Gonzalez}]{Patil23Gorilla}
Shishir~G. Patil, Tianjun Zhang, Xin Wang, and Joseph~E. Gonzalez. 2023.
\newblock \href {https://arxiv.org/abs/2305.15334} {Gorilla: Large language model connected with massive apis}.
\newblock \emph{Preprint}, arXiv:2305.15334.

\bibitem[{Poesia et~al.(2022)Poesia, Polozov, Le, Tiwari, Soares, Meek, and Gulwani}]{Poesia22TST}
Gabriel Poesia, Oleksandr Polozov, Vu~Le, Ashish Tiwari, Gustavo Soares, Christopher Meek, and Sumit Gulwani. 2022.
\newblock \href {https://arxiv.org/abs/2201.11227} {Synchromesh: Reliable code generation from pre-trained language models}.
\newblock \emph{Preprint}, arXiv:2201.11227.

\bibitem[{Schick et~al.(2023)Schick, Dwivedi-Yu, Dessì, Raileanu, Lomeli, Zettlemoyer, Cancedda, and Scialom}]{Schick23Toolformer}
Timo Schick, Jane Dwivedi-Yu, Roberto Dessì, Roberta Raileanu, Maria Lomeli, Luke Zettlemoyer, Nicola Cancedda, and Thomas Scialom. 2023.
\newblock \href {https://arxiv.org/abs/2302.04761} {Toolformer: Language models can teach themselves to use tools}.
\newblock \emph{Preprint}, arXiv:2302.04761.

\bibitem[{Wei et~al.(2023)Wei, Wang, Schuurmans, Bosma, Ichter, Xia, Chi, Le, and Zhou}]{wei2023chainofthought}
Jason Wei, Xuezhi Wang, Dale Schuurmans, Maarten Bosma, Brian Ichter, Fei Xia, Ed~Chi, Quoc Le, and Denny Zhou. 2023.
\newblock \href {https://arxiv.org/abs/2201.11903} {Chain-of-thought prompting elicits reasoning in large language models}.
\newblock \emph{Preprint}, arXiv:2201.11903.

\bibitem[{White et~al.(2023)White, Hays, Fu, Spencer-Smith, and Schmidt}]{White23ChatGPTPrompts}
Jules White, Sam Hays, Quchen Fu, Jesse Spencer-Smith, and Douglas~C. Schmidt. 2023.
\newblock \href {https://arxiv.org/abs/2303.07839} {Chatgpt prompt patterns for improving code quality, refactoring, requirements elicitation, and software design}.
\newblock \emph{Preprint}, arXiv:2303.07839.

\bibitem[{Xu et~al.(2021)Xu, Vasilescu, and Neubig}]{xu2021inide}
Frank~F. Xu, Bogdan Vasilescu, and Graham Neubig. 2021.
\newblock \href {https://arxiv.org/abs/2101.11149} {In-ide code generation from natural language: Promise and challenges}.
\newblock \emph{Preprint}, arXiv:2101.11149.

\end{thebibliography}

\section{Example Appendix}
\label{sec:appendix}

\subsection{Sample with computed Average similarity} \label{ssec:simialrity}

Sample showing how flow similarity is computed for two flows Flow A and Flow B.

\begin{verbatim}

Query = "Post a message in the channel of teams, 
when a new form is created in the forms"

Ground Truth = "triggerOutputs =
await shared_microsoftforms.
CreateFormWebhook({}); 
outputs_shared_teams_PostMessageToConversation
 = shared_teams.PostMessageToConversation(
{ \"poster\": \"User\" });"    

prediction: "triggerOutputs = 
await shared_microsoftforms.
CreateFormWebhook({});
outputs_Get_my_profile_V2 =  
shared_office365users.MyProfile_V2({}); 
outputs_shared_teams_PostMessage
= shared_teams.PostMessageToConversation(
{\"poster\": \"User\",\"location\": 
\"Channel\"});"

API Functions list in ground_truth  = 
[shared_microsoftforms.CreateFormWebhook, 
shared_teams.PostMessageToConversation]


API function list in model generation = 
[shared_microsoftforms.CreateFormWebhook, 
shared_office365users.MyProfile_V2,
shared_teams.PostMessageToConversation]

Similarity Score = 2/3 = 0.666

Since the functions shared_microsoftforms.
CreateFormWebhook and shared_teams.
PostMessageToConversation are found 
in the ground truth.
\end{verbatim}

\subsection{An example of API metdata}
We share a sample of API metadata to highlight the details included in the API description provided to the metaprompt.

\begin{verbatim}

"shared_outlook.SendEmailV2": {
    "FunctionName": "shared_outlook.
                    SendEmailV2",
    "Description": "This operation sends
                     an email message.",
    "IsInTrainingSet": false,
    "DisplayName": "Send an email (V2)",
        "ParametersInfo": [
            {
                "Key": "emailMessage/To",
                "Type": "String",
                "Summary": "To",
                "Format": "email",
                "Description": "Specify 
                    email addresses 
                    separated by semicolons 
                    like someone@contoso.com"
            }, ….
                        ],
        "ResponseSchema": [],
        "IsTrigger": false
    }

\end{verbatim}

\end{document}